\documentclass{iau}
\usepackage{graphicx,natbib}
 
\newcommand{\mnras}{MNRAS}       

\newcommand{\aj}{AJ}

\title{Using 3D Spectroscopy to Probe the Orbital Structure of Composite Bulges}

\author[Erwin et al.]{Peter Erwin$^{1,2}$, Roberto Saglia, Jens Thomas$^{1,2}$,
Maximilian Fabricius$^{1,2}$, Ralf Bender$^{1,2}$, Stephanie Rusli$^{1,2}$, Nina Nowak$^{3}$, 
John E. Beckman$^{4}$, \and Juan Carlos Vega Beltr{\'a}n$^{4}$}

\affiliation{$^1$Max-Planck-Insitut f\"{u}r extraterrestrische Physik, Giessenbachstr., 85748 Garching, Germany \\
$^2$Universit\"{a}ts-Sternwarte M\"{u}nchen, Scheinerstrasse 1,
81679 M\"{u}nchen, Germany \\
$^3$Stockholm University, Department of Astronomy, Oskar Klein Centre, SE-10691 Stockholm, Sweden \\
$^4$Instituto de Astrof\'{\i}sica de Canarias, C/ Via L\'{a}ctea s/n, 38200 La Laguna, Tenerife, Spain}

\pubyear{2014}
\volume{309}
\jname{Galaxies in 3D across the Universe}
\editors{B. L. Ziegler, F. Combes, H. Dannerbauer, M. Verdugo, eds.}

\begin{document}

\maketitle

\begin{abstract}
Detailed imaging and spectroscopic analysis of the centers of nearby S0
and spiral galaxies shows the existence of ``composite bulges'', where
both classical bulges and disky pseudobulges coexist in the same galaxy.
As part of a search for supermassive black holes in nearby galaxy
nuclei, we obtained VLT-SINFONI observations in adaptive-optics mode of
several of these galaxies. Schwarzschild dynamical modeling enables us
to disentangle the stellar orbital structure of the different central
components, and to distinguish the differing contributions of
kinematically hot (classical bulge) and kinematically cool (pseudobulge)
components in the same galaxy.
\end{abstract}

\firstsection
\section{Introduction}

Although the standard picture of the stellar structure of disk galaxies
combines a disk and a central bulge, recent studies have suggested a
dichotomy between galaxies which host \textit{classical} bulges --
round, kinematically hot, and presumed to originate from violent mergers
at high redshift -- and those with \textit{pseudobulges}, where the
central excess stellar light is from a flattened, kinematically cool
structure, presumed to originate from some long-term, internal
(``secular'') processes.

We have recently found evidence that some disk galaxies can harbor both
a classical bulge \textit{and} a disky pseuduobulge (we use the term
``disky pseudobulges'' to distinguish them from bar-derived box/peanut
structures, which are sometimes also called pseudobulges). Evidence for
this includes a combination of highly flattened isophotes, disky
substructures (spirals, nuclear rings, nuclear bars), and stellar
kinematics dominated by rotation in the disky pseudobulge, and rounder
isophotes and stellar kinematics dominated by velocity dispersion in the
classical-bulge region; see \citet{nowak10} and \citet{erwin14} for
details.

As part of our SINFONI Search for Supermassive Black Holes (S$^{3}$BH),
we observed approximately 30 disk and elliptical galaxies with the
SINFONI IFU on the VLT, using natural- or laser-guide-star adaptive
optics to obtain 3D $K$-band spectroscopy of the galaxy centers; our
sample includes three well-defined examples of composite-bulge galaxies.
We combine the  high-resolution stellar kinematics derived from this
data with larger-scale, ground-based spectroscopy and \textit{HST} and
ground-based imaging to measure SMBH masses via Schwarzschild dynamical
modeling \citep[e.g.,][]{nowak07,nowak08,nowak10,rusli11,rusli13}.

\begin{figure}
\centering
\includegraphics[width=1.0\columnwidth]{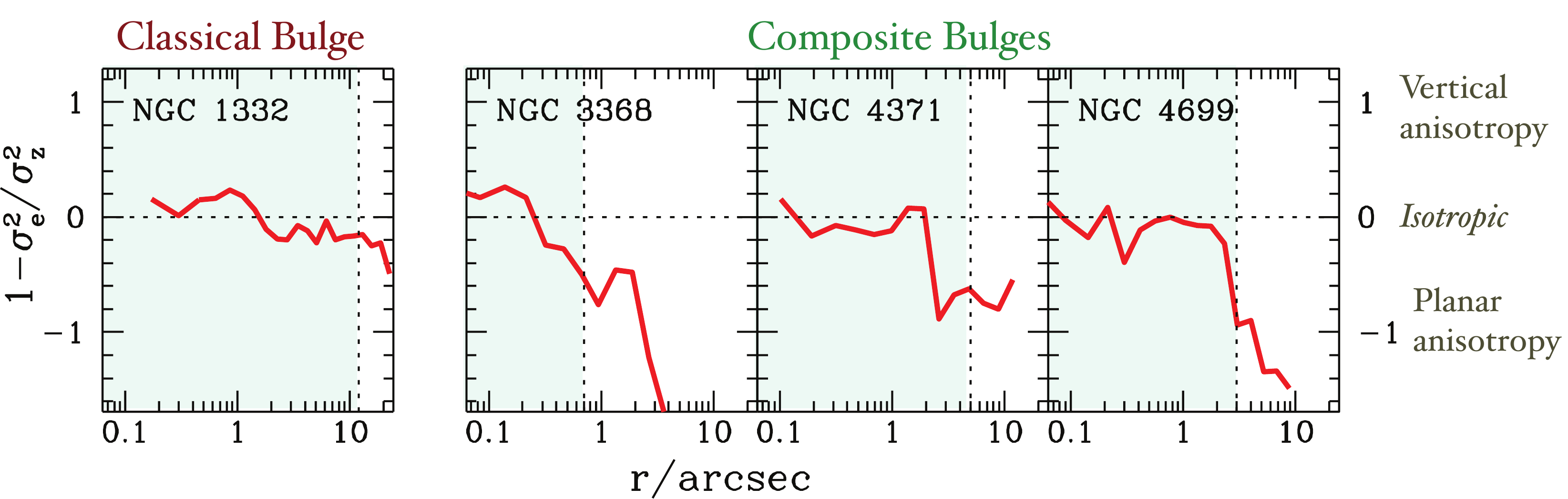} 
\caption{Results of Schwarzschild modeling for classical-bulge S0 galaxy NGC~1332 
\citep[][left]{rusli11} and three composite-bulge galaxies, based on VLT-SINFONI AO data. 
The plots, based on mass-weighted averages of stellar orbits within $\pm23^{\circ}$ of 
the equatorial plane for each galaxy, show equatorial-vs-vertical stellar anisotropy as a function of 
radius; the equatorial term combines radial and tangential dispersions:  
$\sigma_{e}^{2} = (\sigma_{R}^{2} + \sigma_{\phi}^{2})/2$. Vertical dashed lines 
mark the approximate photometric transition between the classical bulge and the disk 
(for NGC~1332) or between the classical bulge and the disky pseudobulge (other three 
galaxies), with the shading indicating the classical-bulge region. For all four galaxies, 
the classical bulge is dominated by isotropic velocity dispersion, while the disk or 
disky pseudobulge regions show planar-dominant anisotropy.}\label{fig1}
\end{figure}

As part of the Schwarzschild modeling process, we obtain weighted
libraries of stellar orbits for the central galaxy regions; these can be
used to explore the relative contributions of ordered (rotational) and
random stellar motions within classical and disky pseudobulge regions.
Fig.~\ref{fig1} shows part of this analysis, plotting the
planar/vertical anisotropy of 3D stellar orbits as a function of radius
for an S0 with a purely classical bulge (NGC~1332) and for three
composite-bulge galaxies. The anisotropy term measures the relative
amounts of ``equatorial'' dispersion (that is, radial and azimuthal
dispersions added in quadrature) versus vertical dispersion (with
respect to the equatorial plane). In all four galaxies, the dispersion
is approximately isotropic within the classical-bulge region, and shifts
to an equatorial-dominant state in the disk outside (for NGC~1332) or in
the disky pseudobulge region (for the composite-bulge galaxies). This is
additional evidence supporting the argument that what we identify as
``classical bulges'' in the composite-bulge systems are isotropic,
pressure-supported components similar to low-luminosity ellipticals,
while the disky pseudobulges have stellar kinematics similar to those of
large-scale disks.

Full details of this study are presented in \citet{erwin14}.

\end{document}